\documentclass[aip,reprint,twocolumn,a4paper,superscriptaddress]{revtex4-1}

\usepackage{tabularx}

\usepackage[version=3]{mhchem} 
\usepackage{graphicx}
\usepackage{xcolor}
\usepackage{siunitx}
\usepackage{textcomp}

\usepackage[flushleft]{threeparttable}
\usepackage{booktabs,multirow}

\usepackage[breaklinks]{hyperref}
\hypersetup{
	pdfnewwindow=true,      
	colorlinks=true,       
	linkcolor=blue,          
	citecolor=blue,        
	filecolor=blue,      
	urlcolor=blue           
}

\begin{document}

\author{Anthony Impellizzeri}
\affiliation{Universit\'e de Nantes, CNRS, Institut des Mat\'eriaux Jean Rouxel, IMN, F-44000 Nantes, France}
\author{Michele Amato}
\email{michele.amato@u-psud.fr}
\affiliation{Universit\'e Paris-Saclay, CNRS, Laboratoire de Physique des Solides, 91405, Orsay, France}
\author{Chris P. Ewels}
\affiliation{Universit\'e de Nantes, CNRS, Institut des Mat\'eriaux Jean Rouxel, IMN, F-44000 Nantes, France}
\author{Alberto Zobelli}
\affiliation{Universit\'e Paris-Saclay, CNRS, Laboratoire de Physique des Solides, 91405, Orsay, France}

\title[Published on J. Phys. Chem. C \textbf{126}, 17746 (2022); \url{http://dx.doi.org/10.1021/acs.jpcc.2c05549}]{Electronic structure of folded hexagonal boron nitride}

\begin{abstract}
Folded regions are commonly encountered in a number of hexagonal boron nitride (h-BN) based bulk and nanostructured materials. Two types of structural modifications occur in folded h-BN layers: local curvature at the folded edges and interlayer shear of the layers which changes the stacking of the overlapping flat regions. In this work we discuss, via density functional theory simulations, the impact of these structural modifications on the ground state electronic structure of the pristine monolayer. We show that, depending on the fold orientation, the overlapping region might present different stacking configurations with subsequent variations of the fundamental band gap; further gap changes occur at the folded regions. The overall electronic structure of a BN folded monolayer can finally be described as a type-II junction between two wide gap semiconductors located at the curved and flat overlapping zones. 
\end{abstract}

\maketitle

\section{Introduction}
Hexagonal boron nitride (h-BN) presents a unique combination of a wide band gap semiconductor and a layered crystal structure.\cite{Cassabois2016a,Arnaud-06,Wirtz2006} These characteristics make it the ideal substrate for 2D materials but they are also at the origin of very original optical properties.\cite{DeanNATURENANOTECH2010} Indeed, while being an indirect band gap semiconductor, h-BN presents a strong luminescence in the far UV, which has been recently interpreted as due to significant exciton-phonon coupling effects.\cite{Cassabois2016a,Cannuccia2019,Paleari2019} Furthermore, point defects in h-BN may act as very bright and room temperature stable single photon sources ranging from the visible to the far UV.\cite{TranNATURENANOTECH2016,Bourrellier2016,Sajid_2020,palla2021optical} The optical response of 2D semiconductors can be easily affected by low energy deformations such as layer gliding, bending or folding. For instance, excitonic emissions at wrinkles in transition metal dichalcogenides are shifted up to tens of meV due to local changes in the electronic band gap associated to strain.\cite{DhakalCM2017} Similar energy shifts have been observed also for single photon sources in bended h-BN.\cite{GrossoNC2017} High curvature regions have a much stronger impact on the high energy h-BN luminescence spectrum: at layer folds the emission spectrum is dominated by near band edge emissions, few hundreds of meV below the free exciton energy.\cite{Bourrellier2014,Schue2DMATER2016} Though it has been demonstrated that these spectral features are linked to phonon assisted inter-valley scattering phenomena,\cite{Cassabois2016} the mechanism behind the selection of specific phonon replica at given folds remains unclear.

Folded regions are very easily encountered in a number of BN based bulk and nanostructured materials. For instance, folds have been directly observed by transmission electron microscopy in flakes obtained by chemical exfoliated~\cite{AlemPRL2012} or chemical vapor deposition~\cite{AnNPJ2019,JangSR2016,SinghAIPADV2019} and, in bulk crystals, at the early stages of the hexagonal to cubic transformation.\cite{CollazoAPL1998,NistorPM2005,HeMRT1998} Furthermore, zigzag edges in BN bilayers reconstruct by forming out-of-plane bonds leading to lobe structures.\cite{AlemPRL2012,XuAFM2017} Similar smooth edge-folded h-BN also occurs in collapsed single-walled boron nitride nanotubes (SWBNNTs), composed of two edge cavities linked via a central flat region similar to bilayer h-BN.\cite{XuAFM2017,TangNANOTECH2006} 
A large fraction of folded BN nanostructures including a significant number of collapsed nanotubes can be obtained through plasma synthesis or the post-synthesis application of high-pressures.\cite{FathalizadehNL2014,Silva-SantosJPCC2021}

Whereas folds are very common in h-BN and their role in defining the ultimate optical properties of defective BN materials has been experimentally proved, their ground state electronic structure has not yet been described. In this work we present first-principles atomistic calculations of folded h-BN layers. 
We discuss changes of the electronic band gap both at the fold and at overlapping regions where bilayer structures with different stacking orders can be obtained. Folded monolayers can finally be described as type II junctions between the curved and flat zones with potential strong implications for the optical response of this class of systems.

\section{Theoretical Methods}
Density Functional Theory (DFT) calculations have been performed using the  AIMPRO package~\cite{RaysonCPC2008,RaysonPRB2009,RAYSONCPC2010} under the generalized gradient approximation (GGA) parametrization by Perdew, Burke, and Ernzerholf (PBE)~\cite{PerdewPRL1996} with D2-London dispersion corrections.\cite{GrimmeJCC2006} The choice of such dispersion scheme is justified by an accurate benchmark structural analysis (see SI, Section 1). Relativistic pseudopotentials from the Hartwigsen-Goedecker-Hütter scheme~\cite{HartwigsenPRB1998} have been employed. Electron wave functions have been expanded in terms of a basis consisting of Gaussian functions multiplied by polynomials including all angular momenta up to \textit{p} ($l = 0-1$) or \textit{d} ($l = 0-2$). Following this nomenclature, a large \textit{pdddp} basis set resulting in 38 independent functions has been used both for boron and nitrogen atoms and a plane wave energy cutoff of 300 Ha for numerical integration. 

A Fermi smearing with an effective temperature of 0.04 eV has been employed for the occupation of the electronic levels. Folded h-BN monolayers have been here modeled by collapsed single walled boron nitride nanotubes (SWBNNT) in order to avoid the presence of any dangling bond. 
Residual interactions between structures in neighboring cells have been reduced by introducing $\sim 2$ nm of vacuum in the non-periodic directions. Atomic positions and unit cell lengths have been simultaneously optimized with no symmetry constraints. The k-point grid for the structural optimization is set to $16 \times 16 \times 12$ for bulk BN while a $16 \times 16 \times 1$ has been chosen in the case the bilayer. For folded monolayers, a $1 \times 1 \times 4$ k-point mesh assured convergence during geometry optimization. 
A high convergence accuracy has been employed for self-consistency ($10^{-7}$ Ha), energy ($10^{-7}$ Ha) and position ($10^{-6}$ $a_{0}$).
Atomic charge distributions have been determined by the Bader charge analysis.\cite{BaderCR1991} Further details of the structural analysis of different hBN stackings for bulk and bilayers are reported in the SI (Section 1).

It is known that semi-local exchange correlation functionals underestimate the quasi-particle energy gaps,\cite{MengleAPLM2019} and for this reason selected systems have been double-checked using the Heyd-Scuseria-Ernzerholf (HSE06) hybrid functional,\cite{HeydJPC2003} as implemented in the plane-waves Quantum-ESPRESSO code.\cite{GiannozziJPCM2017} Hybrid-DFT calculations have been performed by optimizing the set of HSE06 parameters for the specific system considered with respect to state-of-the-art results obtained by many-body perturbation theory (GW). Indeed, as it is known, HSE06 parameters are strongly system-dependent due to the change in screening that geometrical confinement and dimensionality reduction can induce.\cite{DeakPRB2019} The trend obtained via hybrid-DFT calculations is in good agreement with the GGA+D2 description of bulk and bilayers thus confirming the general validity of our results. Further details including benchmark calculations using a range of functionals and dispersion corrections are given in SI, Section 1 and 2.

\section{Results and discussion}
\begin{figure*}[tb]
	\centering
	\includegraphics[width=0.7\textwidth]{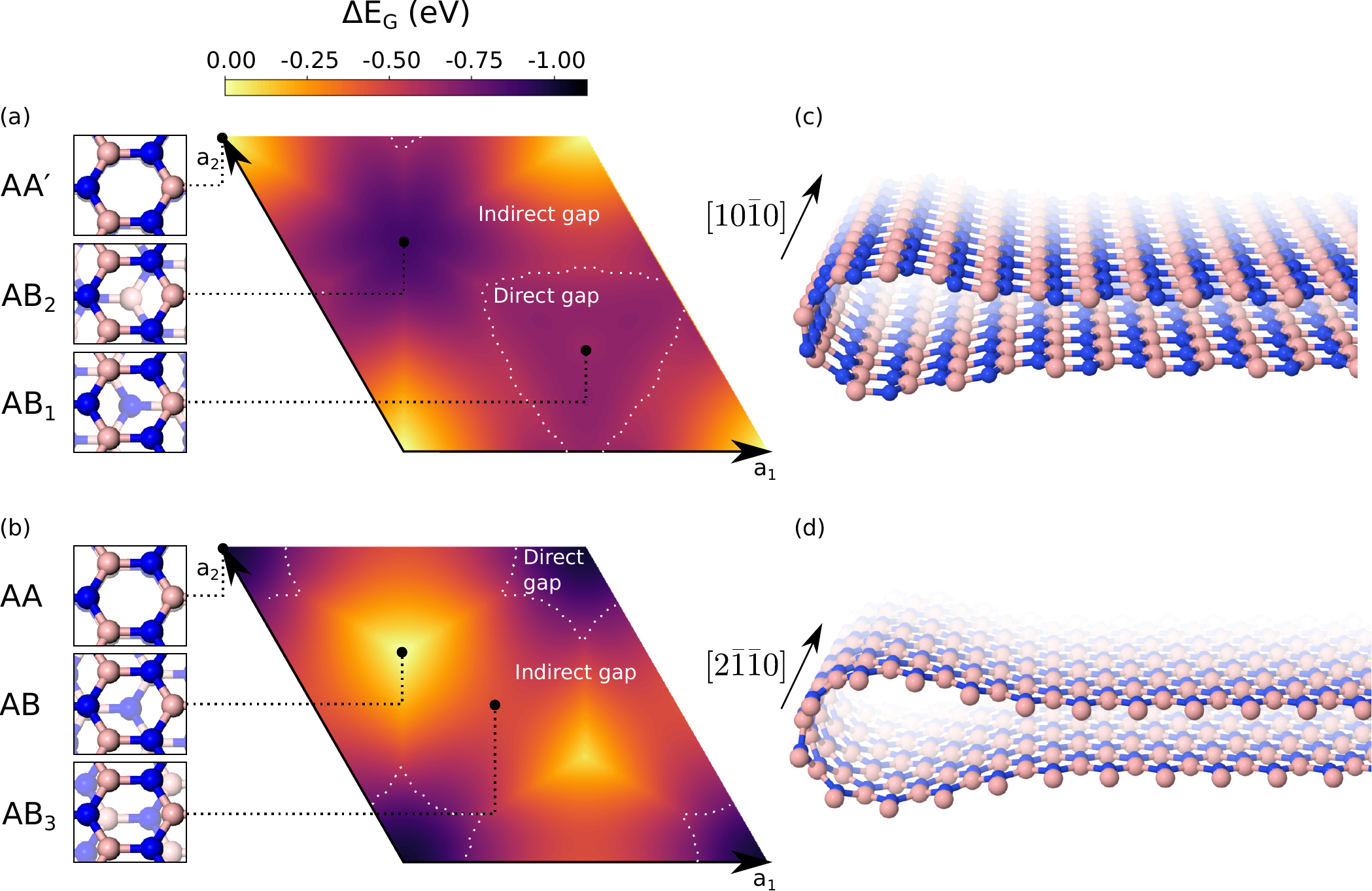}
	\caption{(a,b) Maps of the band gap differences as a function of stackings in BN bilayers. The atomic structures represented in the left side panels correspond to the high symmetry configurations achievable by rigid shifts of the layers. Configurations in (a) can be obtained starting from the AA$^\prime$  stacking while those in (b) from the AA stacking. The zero energy of the maps refers to the most stable configurations namely the AA$^\prime$  in (a) and AB in (b). (c) Atomic structure of an armchair type fold, the flat region can assume only stacking configurations along the major diagonal of map (a). (d) Atomic structure of a zigzag type fold, the flat region can assume only stacking configurations along the minor diagonal of map (b).}
	\label{Fig1}  
\end{figure*}

Two types of structural modifications occur in folded h-BN layers: local curvature at the folded edges and interlayer shear of the layers which changes the stacking of the overlapping flat regions (Figure~\ref{Fig1}.c-d). Both kind of deformations have non-negligible effects on the electronic structure of the folded system. Whereas h-BN has an AA$^\prime$ ground state stacking configuration with boron and nitrogen atoms overlapping in adjacent layers,\cite{PeaseNATURE1950} other non-AA$^\prime$ stacking have also been recently observed in samples synthesized by chemical exfoliation~\cite{WarnerACSNANO2010} and modified vapor deposition.\cite{Gilbert2019} Four high-symmetry stacking configurations can be derived from the AA$^\prime$  stacking through interlayer shearing and in-plane rotations (see Fig.~\ref{Fig1}.a-b). Following a previous nomenclature~\cite{Gilbert2019,Bourrellier2014} we refer to these configurations as AA (homonuclear atoms overlapping in neighbouring layers), AB (analogous to Bernal AB in graphite, with B and N atoms overlapping at the $\beta$ sites), AB$_{1}$ and AB$_{2}$ (as AB but with solely B or N atoms overlapping respectively). A further configuration, at the saddle point between neighboring high symmetry stacking configurations, is here
labelled AB$_{3}$. It is obtained by applying a (0,~1/2,~0) translation to the basal plane of the AA stacking and presents BN dimers sitting at the center of hexagons of the neighbouring plane.

Lattice parameters, interlayer distances, binding energies and band gap obtained at different DFT levels for these high symmetry stackings are summarized in the SI, sections 1 and 2. While GGA provides underestimated values for the electronic band gap, the trends observed for the high symmetry configurations and their relative band gap values are in very good agreement with those obtained both with HSE06 (see SI, Section 2) and recent GW theoretical results.\cite{MengleAPLM2019}
These considerations justify the use of conventional GGA calculations to extend the study of band gap changes to more continuous stacking transformations. Smooth shifts between layers occur for instance at folds or multi-layer wrinkles due to different curvature radii for each layer and they have been extensively discussed in the case of multiwalled BN nanotubes.\cite{Leven2016} In Figure~\ref{Fig1} we report full maps of the band gap variation obtained by continuous translations of one layer with respect to the other starting from an AA$^\prime$ (Figure~\ref{Fig1}.a) and AA bilayer (Figure~\ref{Fig1}.b). The calculated GGA-D2 band gap varies by up to 1 eV with stacking, the most stable configurations showing the largest band gaps. Changes of the registry of the layers can induce a transition from an indirect to a direct band gap. While this electronic transformation has been discussed for high symmetry  bulk and bilayer configurations,\cite{Aggoune-18,MengleAPLM2019} here we generalize this analysis to an arbitrary shearing of the layers and we report in Figure \ref{Fig1}.a,b the domains corresponding to direct and indirect gaps.

The stacking in the flat regions of a folded layer depends from its lattice orientation. When folding along the $[10\bar{1}0]$ direction (Figure~\ref{Fig1}.c), the fold can be named armchair using the nanotube nomenclature. The overlapping flat regions can assume the stacking configurations corresponding to the major diagonal of Figure \ref{Fig1}.a, the most stable of which is the AA$^\prime$ stacking. Folding along the $[2\bar{1}\bar{1}0]$ direction is equivalent to a basal plane rotation of $\pi$. The fold is then zigzag and the stackings achievable correspond to the minor diagonal of Figure~\ref{Fig1}.b, the most stable of which is the AB$_{3}$ (Figure~\ref{Fig1}.d). Whereas this configuration correspond to an energy saddle point, the system is locked to it by the fold direction. This structure is 2.14~meV/atom less stable than the ground state AA$^\prime$ stacking and has a band gap which is about 0.4 eV lower (see SI, Table S3). These results suggest a thermodynamic preference for armchair oriented folds. If a fold is not aligned with a high symmetry direction, the two layers are twisted and a moir\'e or quasi-moir\'e structure appears with a continuous variation of the stacking order. BN moir\'e bilayer structures have been observed for instance in collapsed BN chiral nanotubes by transmission electron microscopy.\cite{FathalizadehNL2014,Silva-SantosJPCC2021} In this case charge localization matching the moir\'e lattice spacing is predicted,\cite{XianNL2019} consistent with the band gap variation shown in Figure~\ref{Fig1}. Explicit calculations of the electronic structure of BN moir\'e bilayers as a function of the twist angle show that the electronic band gap is reduced in respect to the reference AA$^\prime$ structure while keeping an indirect gap nature as discussed in Ref.~\citenum{latil2022}. Therefore, independently from the orientation of the fold in respect to the BN lattice, the overlapping bilayer region will always present a smaller band gap in respect to the unfolded monolayer region. Furthermore, while the monolayer presents a direct band gap, the overlapping region, which could assume the AA$^\prime$,  AB$_{3}$ or a moir\'e stacking configuration, will always present an indirect gap. The overlapping region is therefore expected to act as a funnel for the optical excitation occurring in the direct gap monolayer section.

\begin{figure*}[tb]
	\includegraphics[width=0.75\textwidth]{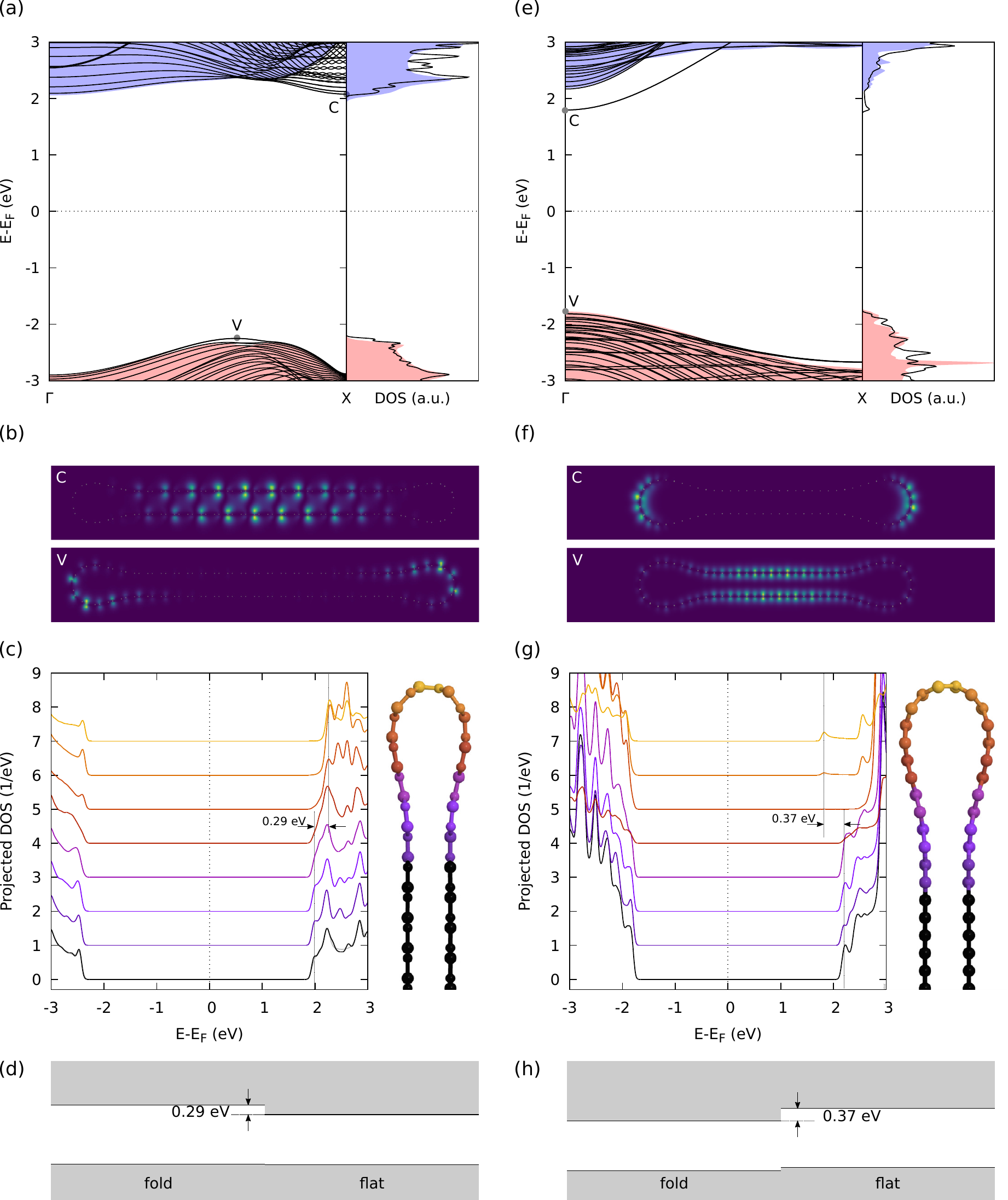}
	\caption{Electronic structure of an armchair (a-d) and a zigzag (e-h) BN fold modeled here via collapsed nanotubes. Electronic bands and density of states of the armchair (a) and zigzag (e) folds (black solid lines) and their respective infinite flat bilayer reference structures AA$^{\prime}$ and AB$_3$ (filled color). (b,f) Electron density map of the valence band maximum V and conduction band minimum C integrated in the direction parallel to the fold. (c,g) Electronic density of states projected at atoms located at different sections of the folds (the colors used for the curve corresponds to the projected atoms in the models aside). 
		(d,h) Schematic representation of the band alignment between folded and flat regions in armchair and zigzag folds.}
	\label{Fig2}      
\end{figure*}    

We now turn to discuss the detailed structure of folded h-BN monolayers which are modeled here by collapsed single walled boron nitride nanotubes (SWBNNT) in order to avoid the presence of any dangling bond. Collapsed BN nanotubes have also an interest in themselves having been directly observed by transmission electron microscopy.\cite{XuAFM2017,TangNANOTECH2006,FathalizadehNL2014,Silva-SantosJPCC2021} As for carbon nanotubes~\cite{HeACSNano2014} SWBNNTs have a critical diameter, above which their circular section collapses to a dogbone-like shape consisting of a central flat region with two side cavities. The collapse occurs when the central attractive interlayer van der Waals interactions overcome the strain-induced energetic cost associated with localizing the tube curvature at the edge cavities; in the case of SWBNNTs we have obtained a critical diameter for this transformation of $\sim 3.05$ nm (smaller than 5.14-5.16~nm for SWCNTs obtained using a similar method,\cite{Impellizzeri2019} due to the higher interlayer binding in BN than graphite).
Here we considered both the cases of armchair folds, modeled by a collapsed (26, 26) armchair SWBNNT (Figure \ref{Fig2}.a-d), and zigzag folds, modeled by a collapsed (38, 0) zigzag SWBNNT (Figure \ref{Fig2}.e-h).
  
The optimized armchair fold presents a central flat region with an AA$^\prime$ -stacking having an interlayer spacing of 0.32~nm. The side cavities have a diameter of 0.54~nm, almost identical to the diameter of a circular (4, 4) nanotube. In Figure~\ref{Fig2}.a we present the fold band structure and total density of states (black curves) overlapped with those of an infinite AA$^\prime$  bilayer (filled regions). The system is characterized by a wide indirect band gap very close to those of the infinite AA$^\prime$  bilayer with a valence band maximum at 2/3 of the $\Gamma$-X path.
Unlike the ideal BN bilayer, the minimum of the conduction band lies at X; this effect can be understood considering that the flat region is an orthorhombic multicell and subsequent band folding effects arise. 

The isodensity surfaces plotted in Figure~\ref{Fig2}.b shows that the valence band maximum is associated with N atoms at the folds, while the conduction band minimum locates at the B atoms of the flat region. In Figure \ref{Fig2}.c we present the density of states projected at atoms located at different regions of the fold. A notable shift of the conduction band edge can be seen when moving towards the fold cavity with an associated increase of the band gap by as much as 0.29 eV. Though tiny (less than 10 meV), a valence band edge upshift from the flat region to the fold can also be observed. The overall structure can then be described as a type II junction between two wide band gap semiconductors: a bulk flat core with a smaller gap and a folded region with a larger one (Figures~\ref{Fig2}.d). 
This representation may need to be slightly revisited when many body effects are taken into account, however the electronic band alignment described here is expected to be preserved. This is because low dimensional systems present lower screening, and the band gap correction will therefore be bigger at the fold than in the flat bilayer region.

The effect of curvature can be analyzed also using Bader charges as a measure of local ionicity. While the central bilayer region shows a 0.72~\ce{e-} transfer from B to N atoms, slightly higher than the equivalent infinite bilayer (0.69~\ce{e-}), in the folds this value drops to 0.56~\ce{e-}, even lower than in an equivalent diameter (4, 4) circular nanotube (0.59~\ce{e-}). This decrease in polarization at the folds corresponds to a destabilization of the bonding caused by the curvature (ionicity is proportional to tube radius in BN nanotubes, see for instance Ref.~\citenum{PhysRevB.68.241405,PhysRevB.67.235406}).

The electronic structure of zigzag folds is presented in Figure \ref{Fig2}.e-h. As discussed above, the most stable stacking configuration for the flat region in a zigzag fold is AB$_{3}$. Compared to the armchair case, zigzag folds have a slightly higher interlayer spacing (0.33~nm) and diameter of the edge cavities (0.63~nm), very close to the diameter of a circular (8, 0) nanotube. Furthermore an additional band appears 0.37~eV below the lowest conduction band (Figure~\ref{Fig2}.e). The origin of this additional band can be mainly attributed to the relative orientation between the tube circumference and consequently to the different B-B next neighbour distances around folded regions between zig-zag and armchair orientation. Indeed, in the zig-zag case they are aligned, maximising the $p_{z}$ orbital overlap between the B atoms, while this is not the case of the armchair orientation. This chirality dependent next-neighbour $p_{z}$-orbital overlap was previously demonstrated in the case of ultra-small diameter nanotubes.\cite{KimPRB2001} This in-gap state remains also when constraining the flat region to the less stable AA stacking (see SI, Figure S3). The minimum of this additional band is associated with the boron atoms at the folded regions and this decreasing of the band gap can be observed also in the projected density of states (Figure~\ref{Fig2}.g). A very tiny valence band energy downshift (less than 10 meV) near the fold region can also be observed. In this case we can hence also describe the folded edge as a type II junction but unlike the armchair folds the larger gap is located in the flat region rather than the folded. Unlike the armchair folds, this picture could be significantly affected when considering many body effects since the band gap is expected to increase more at the fold cavity than at the flat bilayer region. This will lead to a reduction of the conduction band edge mismatch, or even an inversion of the junction order. Although there are several experimental reports of BN armchair folds, zigzag folds have not been reported yet. This can be explained by the high energy penalty of the AB$_3$ stacking compared to the AA$^\prime$ configuration. In addition, armchair folds are obtained from a free zigzag edge whereas zigzag folds correspond to free armchair edges. The latter  structure has a higher formation energy~\cite{gomes2013stability} and for this reason no extended free armchair edges have been observed in BN. 

\section{Conclusions}
In this work we have provided a detailed description of the structure and electronic properties of folded h-BN monolayers. At folded edges two main structural changes occur, a local curvature at the fold and  the appearence of a bilayer structure with a stacking sequence which depends from the orientation of the fold. The overall electronic behavior of the folds can be understood as a combination of these two effects.

Depending on the interlayer stacking, the calculated GGA-D2 band gap of bilayer BN varies by up to 1.15 eV, a trend confirmed via hybrid-DFT calculations using the HSE06 functional. When the fold follows the $[10\bar{1}0]$ direction, the curved region is armchair and the overlapping region can assume the most stable  AA$^\prime$ stacking configuration. Folding along the $[2\bar{1}\bar{1}0]$ direction a zigzag fold is obtained and the most stable configuration in the overlapping region is the AB$_{3}$ stacking. 

In the case of an armchair fold, the curved region presents a larger band gap  in respect to the flat AA$^\prime$  bilayer zone, with an increase of 0.29~eV. In a zigzag fold, the flat AB$_{3}$ stack regions present a gap lower than those of an AA$^\prime$ stacking (with a 0.37~eV reduction) and that can be further reduced in the curved zones.
Therefore, in both cases the overall electronic structure can be described as a type II junction between two high gap semiconductors identified as the fold and the bilayer region. As well as the intrinsic band gap states associated with BN folds demonstrated here, folds are also likely to be preferential locations for point defects such as vacancies,\cite{Zobelli2006}  divacancies~\cite{Zobelli2006,HuJCP2007} and bond-rotation defects.\cite{MoonNANOTECH2004,LiJPC2008} All have been shown to have lower formation energies in regions of high curvature, and all can induce defect-related states in the gap. As such this will only enhance the importance of BN-folding for light emission and absorption.

These ground state electronic effects are expected to have important consequences both for the transport and optical properties of folded BN layers and might provide fundamental insights for understanding the overall physical behavior of defective BN crystals. 

\acknowledgments 

We thank S. Latil and R. Arenal for useful discussions and technical support. M.A. is grateful to D. Varsano for helpful discussions. The work leading to these results received funding from the French Agence Nationale de Recherche Projects ANR-16-CE24-0008-01 EdgeFiller, ANR-20-CE08-0026 OPIFCat and ANR-19-CE30-0007 BONASPES. A. Impellizzeri and C.P. Ewels acknowledge the CCIPL ``Centre de Calcul Intensif Pays de la Loire'' where many of the calculations were performed.

\section*{Supporting information}
Supporting information can be downloaded free of charge at \url{http://dx.doi.org/10.1021/acs.jpcc.2c05549}. Interlayer potentials and structural analysis of bulk and bilayer h-BN; energy gaps of bulk and bilayer h-BN, comparison DFT-LDA, GGA, HSE, GW; electronic structure of AA-stacked collapsed zig-zag h-BN nanotubes.

\bibliography{BNfolds.bib}

\end{document}